\RequirePackage{amsmath,amssymb}
\documentclass{llncs}

\usepackage[utf8]{inputenc}
\usepackage[english]{babel}
\usepackage[ruled]{algorithm2e}
\usepackage{pgfplots}
\usepackage{microtype}
\usepackage{tkz-graph}
	\usetikzlibrary{shapes.geometric}
\usepackage{xcolor}
\usepackage{subfig}
\usepackage{float}
\usepackage{listings}
\usepackage[hidelinks]{hyperref}

\title{Balancing expression dags for more efficient lazy adaptive evaluation}
\author{Martin Wilhelm}
\institute{Institut für Simulation und Graphik, Otto-von-Guericke-Universität~Magdeburg\\
Universitätsplatz 2, D-39106 Magdeburg, Germany}

\newcommand{\gO}[1]{O\left(#1\right)}
\newcommand{\lst}{\text{list}}
\newcommand{\bal}{\text{bal}}
\newcommand{\emax}{e_{\max}}
\newcommand{\RealAlgebraic}{\texttt{Real\_algebraic}}
\newcommand{\p}{q}
\newcommand{\cpp}{C\texttt{++}}
\newcommand{\gpp}{g\texttt{++}}
\newtheorem{observation}{Observation}

\definecolor{SEQCOL1}{RGB}{227,74,51}
\definecolor{SEQCOL2}{RGB}{253,187,132}
\definecolor{SEQCOL3}{RGB}{254,232,200}

\definecolor{QCOL1}{RGB}{241,163,64}
\definecolor{QCOL2}{RGB}{153,142,195}

\DeclareCaptionFormat{cont}{#1 (cont.)#2#3\par}

\begin{document}
\maketitle

\begin{abstract}
Arithmetic expression dags are widely applied in robust geometric computing.
In this paper we restructure expression dags by balancing consecutive additions or multiplications. We predict an asymptotic improvement in running time and experimentally confirm the theoretical results. Finally, we discuss some pitfalls of the approach resulting from changes in evaluation order.
\end{abstract}

\section{Introduction}

\begin{sloppypar}
Most theoretical algorithms in the field of computational geometry are based on the real RAM model of computation and therefore assume exact real number arithmetic at unit cost. Actual processors cannot represent real numbers and instead use floating-point arithmetic.
Computing exact numerical values is expensive and not always possible. Luckily, it is also seldom necessary to compute exact values in order to ensure robustness in geometric algorithms. The Exact Geometric Computation paradigm instead only demands that the decisions made in a geometrical algorithm are correct~\cite{schirra2000,yap1997}.
A common technique used for exact-decisions computation is to store the computation history in an arithmetic expression dag and then adaptively (re)compute the result with a higher precision until a verified decision can be made.
Several expression-dag-based number types have been developed with different evaluation strategies. Strategies can be to gradually increase the precision bottom-up (\texttt{LEA}~\cite{benouamer1993}) or fall back to exact computation (\texttt{CGAL::Lazy\_exact\_nt}~\cite{pion2011}) if a decision cannot be verified, or to use a precision-driven\footnote{This should more correctly be called ``accuracy-driven'', but we use the term ``precision-driven'' throughout this paper for historical reasons.} evaluation (\texttt{leda::real}~\cite{burnikel1996}, \texttt{Core::Expr}~\cite{karamcheti1999,yu2010}, \RealAlgebraic{}~\cite{moerig10}). All of the mentioned number types suffer from high performance overhead compared to standard floating-point arithmetic.

In this work we make an attempt to improve the performance of dag-based number types by restructuring the underlying expression dag in certain situations. Restructuring the expression dag was originally proposed by Yap~\cite{yap1997}. To our knowledge, there is no previous work that actually implements any restructuring strategy. We focus on reducing the depth of an expression dag, i.e.\ the size of the longest path from any node to the root.
The accuracy needed at a node in an expression dag to guarantee a certain error at its root generally increases with the size of the longest path between the node and the root. Therefore a decrease in expression depth can be expected to lead to better error bounds at lower precision and, consequently, to a better performance of dag-based number types.
We restructure the expression dag by ``balancing'' consecutive additions and consecutive multiplications, such that the maximum depth of the involved operators is minimized. By this we also increase the independence of the nodes, which makes it more feasible to parallelize the evaluation. In this work, however, we will not elaborate on advantages regarding parallelization.

We provide a theoretical analysis and evaluate our strategy based on the number type \RealAlgebraic{} introduced by Mörig et al.\ \cite{moerig10}.
\end{sloppypar}

\section{Theoretical Foundation}
\label{sec:theory}

An \emph{expression dag} is a rooted ordered directed acyclic graph, which is either
\begin{enumerate}
	\item A single node containing a number or
	\item A node representing a unary operation $\lbrace \sqrt{}, -\rbrace$ with one, or a binary operation $\lbrace +,-,*,/\rbrace$ with two, not necessarily disjoint, expression dags as children.
\end{enumerate}
We call an expression dag $E'$ whose root is part of another expression dag $E$ a \emph{subexpression of $E$}.

Let $E$ be an expression dag, let $\circ\in\lbrace +,*\rbrace$ and let $E'$ be a subexpression of $E$ with root $r$ of type $\circ$. Let $T$ be a connected subgraph of $E'$, containing $r$, such that all nodes in $T-r$ have at most one predecessor in $E$ and are of type~$\circ$. Then $T$ is a tree and we call $T$ an \emph{operator tree}. The children of the leaves of $T$ in $E$ are called \emph{operands} of $T$.
We restructure $E$ by replacing all maximal operator trees in $E$ by a balanced operator tree with the same number of operands. For a single tree, we call this replacement \emph{balancing the operator tree}. If all maximal operator trees in $E$ are replaced, we call the process \emph{balancing the expression dag}.

We determine the asymptotic running time of a single precision-driven evaluation before and after balancing the expression dag for a series of additions or a series of multiplications. Assumptions on the unit costs for the arithmetic operations and the increase in accuracy are consistent with \texttt{leda::real}, \RealAlgebraic{} and partly with \texttt{Core::Expr}.
\subsection{Addition}
Assume we have a dag-based number type that determines the result of an addition $z=x+y$ with absolute accuracy $\p$ in time $\Theta(\p+\log|z|)$ if $x$ and $y$ are accurate up to $\p+c$ fractional digits, where $c$ is some constant.

Let $x_1,...,x_n$ be distinct floating point numbers with exponent $\leq e, e\geq 0$.
We want to determine the running time to compute $z=\sum_{i=1}^nx_i$ with absolute accuracy $\p$.
Any expression dag for $z$ contains an operator tree consisting of all addition nodes. Assume that the operator tree is a linear list, i.e.\ the computation order is equivalent to \(x_1+(x_2+(x_3+...+(x_{n-1}+x_n)))\). Then the $i$-th addition (counting from the root) must be accurate up to $\p_i=\p+ic$ fractional digits and the magnitude of its result is at most $e_i=e+\lceil\log(n-i)\rceil$.
Therefore we get the time for computing $z$ by adding the time needed on each level as
\begin{align*}
 T_\lst(\p) &= \gO{\sum_{i=0}^{n-1}(\p_i+e_i)} = \gO{\sum_{i=0}^{n-1}(\p+ic+e+\log(n-i))}\\ 
 		   &= \gO{n\p+n^2+ne}
\end{align*}
This bound is tight if all summands have maximum exponent.
Now assume the operator tree is perfectly balanced, i.e.\ the computation order is equivalent to \[((((x_1+x_2)+(x_3+x_4))+...)+(... + ((x_{n-3}+x_{n-2})+(x_{n-1}+x_n))))\] Then at level $i$ there are $2^i$ additions, which must be accurate up to $\p_i=\p+ic$ fractional digits. The magnitude of their result is at most $e_i=e+\log n - i$. So the asymptotic bound for the computation time shrinks to
\[
 T_\bal(\p) = \gO{\sum_{i=0}^{\log n}2^i(\p+ic+e+\log n - i)} = \gO{n\p+n\log n+ne}
\]

\subsection{Multiplication}

For multiplication we assume the number type computes the result of $z=x*y$ with absolute accuracy $\p$ in time $\Theta((\p+\log|z|)^{\log 3})$ if $x$ is accurate up to $\p+c+\lceil\log |y|\rceil$ and $y$ up to $\p+c+\lceil\log |x|\rceil$ fractional digits, where $c$ is some constant. We determine the running time to compute $z=\prod_{i=1}^nx_i$ with absolute accuracy $\p$.

We consider the operator tree consisting of all multiplication nodes in an expression dag for $z$. Let $e\geq 0$ be the maximum exponent of $x_1,...,x_n$. In the unbalanced case the accuracy needed increases by at most $c+e$ with each level top-down, whereas the maximum exponent of the result increases by $e$ bottom-up. Assuming that $x_1,...,x_n$ are exact, we do not need to increase the accuracy of the leaves. Then we get
\begin{align*}
 T_\lst(\p) &= \gO{\sum_{i=0}^{n-1}(\p_i+e_i)^{\log 3}} = \gO{\sum_{i=0}^{n-1}(\p+i(c+e)+(n-i)e)^{\log 3}}\\ 
 		   &= \gO{n\p^{\log 3}+n^{\log 3+1}+n^{\log 3+1}e^{\log 3}}
\end{align*}

This bound is tight if $x_1,...,x_n$ all have exponent $e$. When the operator tree is balanced, the accuracy needed increases by $c+e_{i+1}$ at level $i$, where $e_i=2^{\log n - i}e$, so the requested accuracy at level $i$ is
\[
 \p_i = \p + ic + \sum_{j=0}^{i+1}2^{\log n - j}e \leq \p + ic + 2^{\log n + 1}e
\]
Therefore
\begin{align*}
 T_\bal(\p) &= \gO{\sum_{i=0}^{\log n}2^i(\p + ic + 2^{\log n + 1}e + 2^{\log n - i}e)^{\log 3}}\\
 		   &= \gO{n\p^{\log 3}+n(\log n)^{\log 3}+n^{\log 3+1}e^{\log 3}}
\end{align*}

If $e>0$ the improvement we get from balancing the tree is dominated by the cost for managing the increasing number of integer digits. If one can expect the exponent to be bounded from above, the improvement gets asymptotically significant.
Let $\emax$ be the largest exponent occuring during the whole computation. Then
\begin{align*}
 T_\lst(\p) &= \gO{\sum_{i=0}^{n-1}(\p+i(c+\emax)+\emax)^{\log 3}}\\
 		    &= \gO{n\p^{\log 3}+n^{\log 3+1}+n^{\log 3+1}\emax^{\log 3}}
\end{align*}

whereas
\begin{align*}
 T_\bal(\p) &= \gO{\sum_{i=0}^{\log n}2^i(\p + i(c+\emax) + \emax)^{\log 3}}\\
 		   &= \gO{n\p^{\log 3}+n(\log n)^{\log 3}+n(\log n)^{\log 3}\emax^{\log 3}}
\end{align*}

The asymptotic bound for $T_\lst(\p)$ is tight if the values of $\Theta(n)$ inner nodes are of order $\emax$.

\section{Implementation}
\label{sec:implementation}

The balancing strategy has been implemented for the dag-based exact-decision numbertype \RealAlgebraic{} designed by Mörig et al \cite{moerig10}. This number type consists of a single- or multi-layer floating-point-filter~\cite{fortune1993}, which falls back to adaptive evaluation with bigfloats stored in an expression dag~\cite{dube93}. Balancing is done at most once at each node, right before the first bigfloat evaluation. Otherwise, existing results would have to be recomputed after changing the structure of the dag, which could potentially lead to a massive overhead if subexpressions need to be evaluated during dag construction. Once evaluated nodes are therefore treated as operands in any subsequent balancing process.

We call evaluations of subexpressions of an expression dag $E$ \emph{partial evaluations of $E$}. By preventing evaluated nodes to be part of another operator tree, frequent partial evaluations during construction can fully negate the benefits of the balancing strategy. If partial evaluations occur only sporadically, then their impact on the \emph{expression depth of $E$}, i.e.\ the maximum distance of any node to its root, is small, since each of the involved subexpressions have been balanced when they were evaluated for the first time.

\begin{observation}
\label{obs:partial}
Let $E$ be an expression dag consisting of $n$ additions or $n$ multiplications and $n+1$ bigfloats. Let $d$ be the expression depth of $E$ after its evaluation. If at most $k$ partial evaluations of $E$ occur before the evaluation of $E$, then $d \leq k\lceil\log \frac{n}{k}\rceil$. 
\end{observation}

On the first evaluation of a node that cannot be handled by the floating-point-filter, the balancing process starts at this node recursively. If the current node contains an addition or a multiplication and has not be balanced before, all operands of the maximal operator tree containing this node as root will be retrieved. If the depth of the operator tree can be reduced, it gets balanced (cf.\ Algorithm~\ref{alg:retrievetrees}). For almost-balanced trees a slight decrease in depth may not justify restructuring a large tree. Therefore it might be useful to experimentally decide on a factor tightening this condition in later implementations.

\begin{algorithm}[h]
 \KwData{current node $node$}
 \If{$node$ is not balanced}{
   \eIf{$node$ is addition or multiplication}{
 	  $(operands, depth) = retrieve\_operands(node)$
 	
 	  \If {$depth > \lceil \log |operands| \rceil$} {
 		  balance current operation
 	  }
 	  \ForEach {$op\in operands$}{
 		  recurse on $op$
 	  }
   }{
 	  recurse on children
   }
   mark $node$ as balanced
 }
\caption{The relevant operator trees are retrieved in the form of an operand list with an associated depth. By comparing the depth with the number of operands it gets decided whether the trees should be balanced.} 
\label{alg:retrievetrees}
\end{algorithm}

The operands get retrieved through a depth-first search. Nodes can be retrieved more than once and therefore the same node can represent multiple operands. A node is treated as an operand if one of the following conditions holds:
\begin{enumerate}
 \item The node is not of the same type as the current operation (i.e.\ $+$ or $*$).
 \item The node has already been balanced (and therefore initialized).
 \item The node has more than one parent.
\end{enumerate}

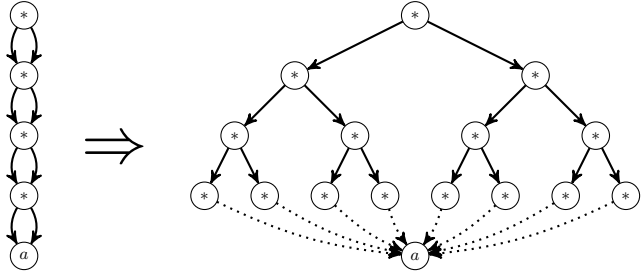
\begin{figure}[h]
\centering
\begin{tikzpicture}
\begin{scope}[scale=0.4,shift={(-13,0)}]
  \SetGraphUnit{2} 
  \tikzstyle{VertexStyle} = [scale=0.7,circle,draw]

   \Vertex[Math,L=*]{P4} 
   \SO[Math,L=*](P4){P3}
   \SO[Math,L=*](P3){P2}
   \SO[Math,L=*](P2){P1}
   \SO[Math,L=a](P1){A}
   
  \tikzstyle{EdgeStyle}   = [->,>=stealth',bend left]
   \Edges(P4,P3,P2,P1,A)
   
  \tikzstyle{EdgeStyle}   = [->,>=stealth',bend right]
   \Edges(P4,P3,P2,P1,A)

\end{scope}
\begin{scope}[shift={(-4,-1.8)}]
\node{\Huge $\Rightarrow$};
\end{scope}
\begin{scope}[scale=0.4]
  \tikzstyle{VertexStyle} = [scale=0.7,circle,draw]

   \Vertex[Math,L=*]{P0} 
   
  \tikzstyle{VertexStyle} = []
   \SOWE[NoLabel](P0){P11P} \SOEA[NoLabel](P0){P12P}
   \WE[NoLabel](P11P){P11PP} \EA[NoLabel](P12P){P12PP}
   \WE[NoLabel](P11PP){P11PPP} \EA[NoLabel](P12PP){P12PPP}
   
  \tikzstyle{VertexStyle} = [scale=0.7,circle,draw]
   \SOWE[Math,L=*](P11PPP){P11} \SOEA[Math,L=*](P12PPP){P12}
   
  \tikzstyle{VertexStyle} = []
   \SOWE[NoLabel](P11){P21P} \SOEA[NoLabel](P11){P22P}
   \SOWE[NoLabel](P12){P23P} \SOEA[NoLabel](P12){P24P}
   
  \tikzstyle{VertexStyle} = [scale=0.7,circle,draw]
   \SOWE[Math,L=*](P21P){P21} \SOEA[Math,L=*](P22P){P22}
   \SOWE[Math,L=*](P23P){P23} \SOEA[Math,L=*](P24P){P24}
   
  \tikzstyle{VertexStyle} = []
   \SOWE[NoLabel](P21){P31P} \SOEA[NoLabel](P21){P32P}
   \SOWE[NoLabel](P22){P33P} \SOEA[NoLabel](P22){P34P}
   \SOWE[NoLabel](P23){P35P} \SOEA[NoLabel](P23){P36P}
   \SOWE[NoLabel](P24){P37P} \SOEA[NoLabel](P24){P38P}
   
  \tikzstyle{VertexStyle} = [scale=0.7,circle,draw]
   \SO[Math,L=*](P31P){P31} \SO[Math,L=*](P32P){P32}
   \SO[Math,L=*](P33P){P33} \SO[Math,L=*](P34P){P34}
   \SO[Math,L=*](P35P){P35} \SO[Math,L=*](P36P){P36}
   \SO[Math,L=*](P37P){P37} \SO[Math,L=*](P38P){P38}
   
  \tikzstyle{VertexStyle} = []
   \SOEA[NoLabel](P34){AP}
   
  \tikzstyle{VertexStyle} = [scale=0.7,circle,draw]
   \SO[Math,L=a](AP){A}
   
  \tikzstyle{EdgeStyle}   = [->,>=stealth']
   \Edges(P0,P11,P21,P31)
   \Edges(P21,P32)
   \Edges(P11,P22,P33)
   \Edges(P22,P34)
   \Edges(P0,P12,P23,P35)
   \Edges(P23,P36)
   \Edges(P12,P24,P37)
   \Edges(P24,P38)
  
  \tikzstyle{EdgeStyle}   = [->,>=stealth',dotted,bend right=10]
   \Edges(P31,A)
   \Edges(P32,A)
   \Edges(P33,A)
   \Edges(P34,A)
  \tikzstyle{EdgeStyle}   = [->,>=stealth',dotted,bend left=10]
   \Edges(P35,A)
   \Edges(P36,A)
   \Edges(P37,A)
   \Edges(P38,A)
\end{scope}
\end{tikzpicture}
\caption{An expression dag computing $a^{16}$ and its exponential expansion that results from resolving multiple references through copying.}
\label{img:exponentialblowup}
\end{figure}

The third condition is necessary, since the subexpression represented by this node will be destroyed during balancing. If a full copy of the node would be created instead, this may lead to an exponential blowup for highly self-referential structures (cf.\ Figure~\ref{img:exponentialblowup}).

A similar observation as for the second condition (cf.\ Observation~\ref{obs:partial}) can be made. If few operator nodes have more than one parent, the overall impact on the expression depth is small.

\begin{observation}
Let $E$ be an expression dag consisting of $n$ additions or $n$ multiplications and $n+1$ bigfloats. Let $d$ be the expression depth of $E$ after its evaluation. If at most $k$ operator nodes of $E$ have more than one reference, then $d \leq k\lceil\log \frac{n}{k}\rceil$. 
\end{observation}

We balance an operation by combining two operands to a new operand until only one node is left by treating the operand vector like a queue (cf.\ Algorithm~\ref{alg:balanceoperation}). Note that this strategy does not preserve the evaluation order of the operands if the number of operands is not a power of two. This can have consequences for the running time and may obfuscate the experiments. If operand order is of importance, it can be preserved by inserting dummy nodes with values $0$ for addition and $1$ for multiplication up to the next power of two.

\begin{algorithm}[h]
 \KwData{operand vector $operands$, operation type $\circ$, root node $root$}
 $size = operands.size()$\;
 \For{$i=0$ \textbf{to} $size-2$}{
   $operands.add(new\ Node(operands[2i],operands[2i+1],\circ))$
 }
 $root.left = operands[2*size-2]$\;
 $root.right = operands[2*size-1]$\;
\caption{The operator tree is restructured by discarding all nodes except the root and building a new balanced operator tree bottom-up by repeatedly combining the two smallest subtrees to a new tree.} 
\label{alg:balanceoperation}
\end{algorithm}

\section{Experiments}

All experiments are run on an Intel Core i5 660 with 8GB RAM under Ubuntu 16.04 LTS. We use Boost interval arithmetic as floating-point-filter and MPFR bigfloats for the bigfloat arithmetic. The code is compiled using \gpp~5.4.0 with \cpp11 on optimization level \texttt{O3} and linked against Boost~1.62.0 and MPFR~3.1.0. Test results are averaged over 25 runs each if not stated otherwise. The variance for each data point is negligible. 

We will perform two simple experiments to evaluate our strategy. In our first experiment we compute the sum of the square roots of the natural numbers $1$ to $n$ with accuracy $\p$.

\begin{lstlisting}[escapechar=\%]
template <class NT> void sum_of_sqrts(const int n, const long %\p%){
  NT sum = NT(0);
  for (int i = 1; i <= n; ++i) {
    sum += sqrt(NT(i));
  }
  sum.guarantee_absolute_error_two_to(%\p%);
}
\end{lstlisting}

The second test computes the generalized binomial coefficient \begin{center}\({\sqrt{13}\choose n} = \frac{\sqrt{13}(\sqrt{13}-1)\cdots(\sqrt{13}-n+1)}{n(n-1)\cdots 1}\)\end{center} with accuracy $\p$.

\begin{lstlisting}[escapechar=\%]
template <class NT> void bin_coeff(const int n, const long %\p%){
  NT b = sqrt(NT(13));
  NT num = NT(1); NT denom = NT(1);
  for (int i = 0; i < n; ++i) {
    num *= b - NT(i);
    denom *= NT(i+1);
  }
  NT bc = num/denom;
  bc.guarantee_absolute_error_two_to(%\p%);
}
\end{lstlisting}

For each test we compare four different implementations. We distinguish between no balancing (def), balancing only addition (add), balancing only multiplication (mul) and balancing both addition and multiplication (all).

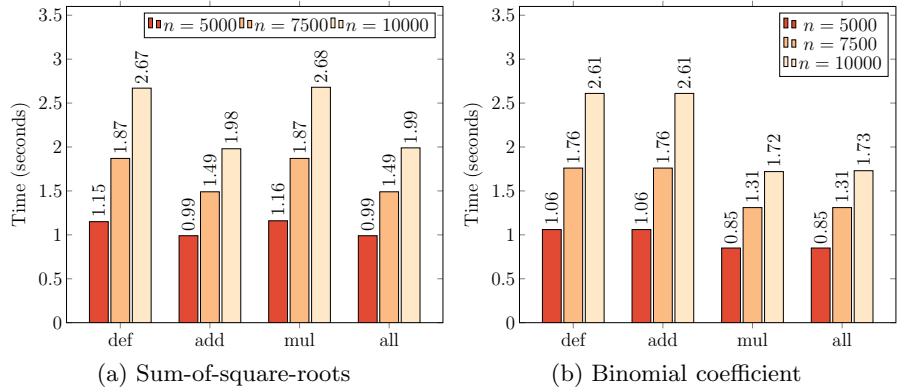
\begin{figure}[htb]
\centering
\subfloat[Sum-of-square-roots]{
\begin{tikzpicture}[scale=0.5]
\begin{axis}[
	symbolic x coords={def,add,mul,all},
	xtick=data,
	enlarge x limits=0.2,
	ylabel=Time (seconds),
	ymin = 0,
	ymax = 3.6,
    width  = 0.95*\textwidth,
    bar width = 0.5cm,
    ybar,
    label style={font=\Large},
    tick label style={font=\Large},
    legend style={font=\Large},
    legend columns=-1,
    nodes near coords,
    every node near coord/.append style={rotate=90, anchor=west, 
    									 /pgf/number format/precision=4,font=\Large}
]

\addplot[fill=SEQCOL1] coordinates {(def,1.15) (add,0.99) (mul,1.16) (all,0.99) };
\addplot[fill=SEQCOL2] coordinates {(def,1.87) (add,1.49) (mul,1.87) (all,1.49) };
\addplot[fill=SEQCOL3] coordinates {(def,2.67) (add,1.98) (mul,2.68) (all,1.99) };
	
\legend{$n=5000$,$n=7500$,$n=10000$}
\end{axis}
\end{tikzpicture}
}
\subfloat[Binomial coefficient]{
\begin{tikzpicture}[scale=0.5]
\begin{axis}[
	symbolic x coords={def,add,mul,all},
	xtick=data,
	enlarge x limits=0.2,
	ylabel=Time (seconds),
	ymin = 0,
	ymax = 3.6,
    width  = 0.95*\textwidth,
    bar width = 0.5cm,
    ybar,
    label style={font=\Large},
    tick label style={font=\Large},
    legend style={font=\Large},
    nodes near coords,
    every node near coord/.append style={rotate=90, anchor=west, 
    									 /pgf/number format/precision=4,font=\Large}
]

\addplot[fill=SEQCOL1] coordinates {(def,1.06) (add,1.06) (mul,0.85) (all,0.85) };
\addplot[fill=SEQCOL2] coordinates {(def,1.76) (add,1.76) (mul,1.31) (all,1.31) };
\addplot[fill=SEQCOL3] coordinates {(def,2.61) (add,2.61) (mul,1.72) (all,1.73) };
	
\legend{$n=5000$,$n=7500$,$n=10000$}
\end{axis}
\end{tikzpicture}
}
\caption{Performance gain through balancing for \texttt{sum\_of\_sqrts} and \texttt{bin\_coeff} with a requested accuracy of $\p=50000$ for different values of $n$.}
\label{fig:sosrbc50}
\end{figure}

The sum-of-square-roots test as well as the binomial coefficient test provide simple examples for when balancing can be of use (cf.\ Figure~\ref{fig:sosrbc50}). Obviously balancing multiplication does not have a positive effect on the sum-of-square-roots test and balancing addition does not have a positive effect on the binomial coefficient computation. There is a small overhead in these cases due to the traversal of the dag. The overhead vanishes in \texttt{all}, since the same procedure is used for both addition and multiplication.

\begin{figure}[htb]
\centering
\subfloat[Sum-of-square-roots]{
\begin{tikzpicture}[scale=0.5]
\begin{axis}[
	symbolic x coords={def,add,mul,all},
	xtick=data,
	enlarge x limits=0.2,
	ylabel=Time (seconds),
	ymin = 0,
	ymax = 1.7,
    width  = 0.95\textwidth,
    bar width = 0.5cm,
    ybar,
    label style={font=\Large},
    tick label style={font=\Large},
    legend style={font=\Large},
    legend columns=-1,
    nodes near coords,
    every node near coord/.append style={rotate=90, anchor=west, 
    									 /pgf/number format/precision=4,font=\Large}
]
\addplot[fill=SEQCOL1] coordinates {(def,0.44) (add,0.33) (mul,0.44) (all,0.33) };
\addplot[fill=SEQCOL2] coordinates {(def,0.75) (add,0.49) (mul,0.76) (all,0.50) };
\addplot[fill=SEQCOL3] coordinates {(def,1.13) (add,0.66) (mul,1.15) (all,0.66) };
	
\legend{$n=5000$,$n=7500$,$n=10000$}
\end{axis}
\end{tikzpicture}
}
\subfloat[Binomial coefficient]{
\begin{tikzpicture}[scale=0.5]
\begin{axis}[
	symbolic x coords={def,add,mul,all},
	xtick=data,
	enlarge x limits=0.2,
	ylabel=Time (seconds),
	ymin = 0,
	ymax = 1.7,
    width  = 0.95\textwidth,
    bar width = 0.5cm,
    ybar,
    label style={font=\Large},
    tick label style={font=\Large},
    legend style={font=\Large},
    nodes near coords,
    every node near coord/.append style={rotate=90, anchor=west, 
    									 /pgf/number format/precision=4,font=\Large}
]
\addplot[fill=SEQCOL1] coordinates {(def,0.51) (add,0.51) (mul,0.34) (all,0.34) };
\addplot[fill=SEQCOL2] coordinates {(def,0.89) (add,0.90) (mul,0.51) (all,0.51) };
\addplot[fill=SEQCOL3] coordinates {(def,1.40) (add,1.40) (mul,0.68) (all,0.68) };
	
\legend{$n=5000$,$n=7500$,$n=10000$}
\end{axis}
\end{tikzpicture}
}
\caption{Performance gain through balancing for \texttt{sum\_of\_sqrts} and \texttt{bin\_coeff} with a requested accuracy of $\p=25000$. The relative gain is larger than for $\p=50000$ (cf.\ Figure~\ref{fig:sosrbc50}).}
\label{fig:sosrbc25}
\end{figure}
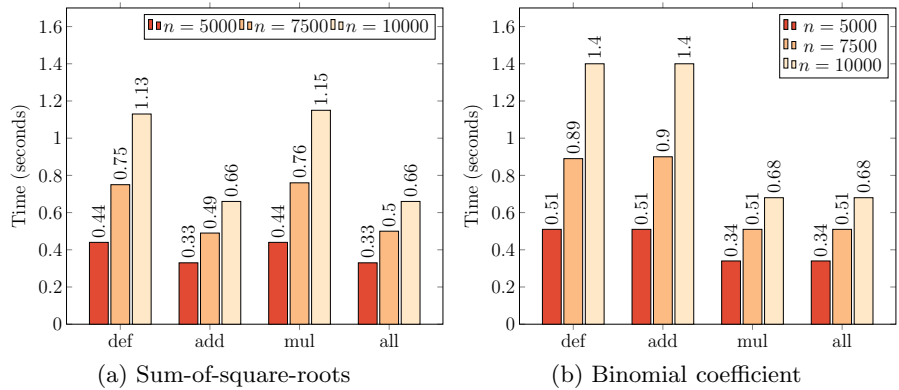

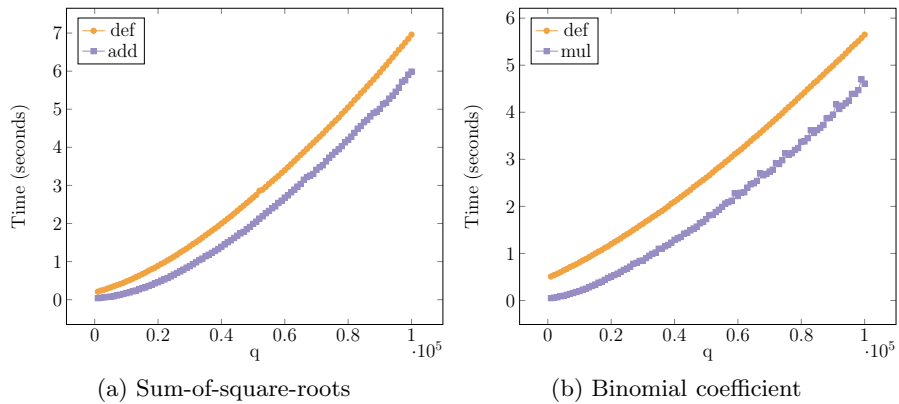
\begin{figure}[!b]
\centering
\subfloat[Sum-of-square-roots]{
\begin{tikzpicture}[scale=0.5]
\begin{axis}[
	xlabel=\p,
	xtick={0,20000,40000,60000,80000,100000},
	ylabel=Time (seconds),
    label style={font=\Large},
    tick label style={font=\Large},
    legend style={font=\Large},
    legend pos=north west,
    width  = 0.95*\textwidth
]
\addplot[QCOL1,mark=*] coordinates { (1000,0.21) (2000,0.24) (3000,0.26) (4000,0.29) (5000,0.32) (6000,0.35) (7000,0.38) (8000,0.41) (9000,0.44) (10000,0.48) (11000,0.51) (12000,0.55) (13000,0.59) (14000,0.63) (15000,0.67) (16000,0.71) (17000,0.76) (18000,0.8) (19000,0.84) (20000,0.89) (21000,0.94) (22000,0.98) (23000,1.03) (24000,1.08) (25000,1.13) (26000,1.18) (27000,1.24) (28000,1.29) (29000,1.34) (30000,1.4) (31000,1.46) (32000,1.51) (33000,1.57) (34000,1.63) (35000,1.69) (36000,1.75) (37000,1.81) (38000,1.87) (39000,1.93) (40000,2) (41000,2.06) (42000,2.13) (43000,2.19) (44000,2.26) (45000,2.33) (46000,2.4) (47000,2.47) (48000,2.54) (49000,2.61) (50000,2.68) (51000,2.75) (52000,2.86) (53000,2.89) (54000,2.96) (55000,3.04) (56000,3.11) (57000,3.18) (58000,3.25) (59000,3.33) (60000,3.4) (61000,3.48) (62000,3.56) (63000,3.64) (64000,3.72) (65000,3.8) (66000,3.88) (67000,3.96) (68000,4.04) (69000,4.12) (70000,4.2) (71000,4.28) (72000,4.36) (73000,4.45) (74000,4.53) (75000,4.62) (76000,4.7) (77000,4.78) (78000,4.88) (79000,4.96) (80000,5.05) (81000,5.14) (82000,5.24) (83000,5.32) (84000,5.42) (85000,5.5) (86000,5.6) (87000,5.69) (88000,5.78) (89000,5.87) (90000,5.97) (91000,6.07) (92000,6.16) (93000,6.26) (94000,6.36) (95000,6.46) (96000,6.56) (97000,6.65) (98000,6.75) (99000,6.85) (100000,6.96)};

\addplot[QCOL2,mark=square*] coordinates { (1000,0.04) (2000,0.05) (3000,0.06) (4000,0.07) (5000,0.08) (6000,0.09) (7000,0.11) (8000,0.13) (9000,0.15) (10000,0.17) (11000,0.19) (12000,0.22) (13000,0.24) (14000,0.27) (15000,0.3) (16000,0.32) (17000,0.36) (18000,0.39) (19000,0.43) (20000,0.46) (21000,0.5) (22000,0.53) (23000,0.57) (24000,0.61) (25000,0.66) (26000,0.7) (27000,0.74) (28000,0.79) (29000,0.84) (30000,0.88) (31000,0.93) (32000,0.98) (33000,1.04) (34000,1.08) (35000,1.14) (36000,1.18) (37000,1.23) (38000,1.28) (39000,1.33) (40000,1.39) (41000,1.45) (42000,1.51) (43000,1.56) (44000,1.62) (45000,1.68) (46000,1.74) (47000,1.78) (48000,1.85) (49000,1.91) (50000,1.99) (51000,2.05) (52000,2.13) (53000,2.19) (54000,2.27) (55000,2.33) (56000,2.4) (57000,2.45) (58000,2.54) (59000,2.6) (60000,2.68) (61000,2.74) (62000,2.82) (63000,2.89) (64000,2.97) (65000,3.04) (66000,3.15) (67000,3.22) (68000,3.26) (69000,3.3) (70000,3.41) (71000,3.48) (72000,3.53) (73000,3.63) (74000,3.73) (75000,3.8) (76000,3.88) (77000,3.95) (78000,4.03) (79000,4.13) (80000,4.2) (81000,4.28) (82000,4.38) (83000,4.49) (84000,4.56) (85000,4.66) (86000,4.72) (87000,4.81) (88000,4.9) (89000,4.94) (90000,5.01) (91000,5.13) (92000,5.17) (93000,5.28) (94000,5.36) (95000,5.46) (96000,5.57) (97000,5.72) (98000,5.76) (99000,5.91) (100000,5.99)};
	
\legend{def,add}
\end{axis}
\end{tikzpicture}
}
\subfloat[Binomial coefficient]{
\begin{tikzpicture}[scale=0.5]
\begin{axis}[
	xlabel=\p,
	xtick={0,20000,40000,60000,80000,100000},
	ylabel=Time (seconds),
    label style={font=\Large},
    tick label style={font=\Large},
    legend style={font=\Large},
    legend pos=north west,
    width  = 0.95*\textwidth
]
\addplot[QCOL1,mark=*] coordinates { (1000,0.51) (2000,0.54) (3000,0.57) (4000,0.6) (5000,0.64) (6000,0.67) (7000,0.71) (8000,0.74) (9000,0.77) (10000,0.81) (11000,0.85) (12000,0.88) (13000,0.92) (14000,0.96) (15000,1) (16000,1.04) (17000,1.07) (18000,1.12) (19000,1.15) (20000,1.2) (21000,1.24) (22000,1.28) (23000,1.32) (24000,1.37) (25000,1.41) (26000,1.45) (27000,1.5) (28000,1.54) (29000,1.59) (30000,1.63) (31000,1.68) (32000,1.72) (33000,1.77) (34000,1.81) (35000,1.86) (36000,1.9) (37000,1.95) (38000,2) (39000,2.06) (40000,2.1) (41000,2.15) (42000,2.2) (43000,2.25) (44000,2.3) (45000,2.35) (46000,2.41) (47000,2.46) (48000,2.51) (49000,2.56) (50000,2.61) (51000,2.66) (52000,2.72) (53000,2.78) (54000,2.83) (55000,2.88) (56000,2.94) (57000,2.99) (58000,3.05) (59000,3.11) (60000,3.16) (61000,3.21) (62000,3.27) (63000,3.33) (64000,3.39) (65000,3.45) (66000,3.5) (67000,3.56) (68000,3.62) (69000,3.68) (70000,3.74) (71000,3.8) (72000,3.86) (73000,3.92) (74000,3.98) (75000,4.05) (76000,4.11) (77000,4.17) (78000,4.23) (79000,4.3) (80000,4.36) (81000,4.42) (82000,4.49) (83000,4.55) (84000,4.62) (85000,4.67) (86000,4.74) (87000,4.8) (88000,4.87) (89000,4.93) (90000,4.99) (91000,5.06) (92000,5.12) (93000,5.18) (94000,5.25) (95000,5.32) (96000,5.39) (97000,5.45) (98000,5.51) (99000,5.58) (100000,5.65)};

\addplot[QCOL2,mark=square*] coordinates { (1000,0.05) (2000,0.06) (3000,0.07) (4000,0.09) (5000,0.1) (6000,0.12) (7000,0.14) (8000,0.16) (9000,0.18) (10000,0.2) (11000,0.23) (12000,0.25) (13000,0.28) (14000,0.31) (15000,0.34) (16000,0.38) (17000,0.4) (18000,0.44) (19000,0.48) (20000,0.51) (21000,0.54) (22000,0.57) (23000,0.61) (24000,0.65) (25000,0.68) (26000,0.72) (27000,0.78) (28000,0.8) (29000,0.84) (30000,0.85) (31000,0.9) (32000,0.95) (33000,0.99) (34000,1.02) (35000,1.1) (36000,1.1) (37000,1.15) (38000,1.18) (39000,1.23) (40000,1.29) (41000,1.32) (42000,1.35) (43000,1.42) (44000,1.44) (45000,1.49) (46000,1.53) (47000,1.57) (48000,1.64) (49000,1.67) (50000,1.73) (51000,1.81) (52000,1.82) (53000,1.88) (54000,1.93) (55000,1.99) (56000,2.06) (57000,2.09) (58000,2.12) (59000,2.28) (60000,2.22) (61000,2.29) (62000,2.3) (63000,2.4) (64000,2.47) (65000,2.49) (66000,2.54) (67000,2.7) (68000,2.66) (69000,2.69) (70000,2.74) (71000,2.78) (72000,2.92) (73000,2.9) (74000,2.98) (75000,3.13) (76000,3.1) (77000,3.13) (78000,3.2) (79000,3.24) (80000,3.37) (81000,3.38) (82000,3.45) (83000,3.62) (84000,3.56) (85000,3.62) (86000,3.67) (87000,3.73) (88000,3.87) (89000,3.88) (90000,3.95) (91000,4.17) (92000,4.07) (93000,4.14) (94000,4.19) (95000,4.25) (96000,4.39) (97000,4.39) (98000,4.47) (99000,4.7) (100000,4.61)};

\legend{def,mul}
\end{axis}
\end{tikzpicture}
}
\caption{Absolute performance gain through balancing for \texttt{sum\_of\_sqrts} and \texttt{bin\_coeff} for $n=10000$ with different requested accuracies (average over five runs). The absolute gain is almost independent of $\p$. The relative gain decreases.}
\label{fig:sosrbcseries}
\end{figure}
The relative benefit of balancing increases if the precision increase due to the number of operands is large relative to the requested accuracy for the result. Figure~\ref{fig:sosrbc25} shows the performance gain through balancing for a requested accuracy of $\p=25000$. With $10000$ operands, the relative gain is about $42\%$ for \texttt{sum\_of\_sqrts} and $51\%$ for \texttt{bin\_coeff} compared to $26\%$ and $34\%$ for $\p=50000$. 
The theoretical analysis from Section~\ref{sec:theory} predicts that the absolute performance gain primarily depends on the number of the operands that can be balanced and is independent from the requested accuracy. The experimental results largely confirm this assumption as shown in Figure~\ref{fig:sosrbcseries}. Since balancing is done before the first evaluation, the overhead due to the balancing procedure only depends on the size of the expression dag and the number of operands.

\section{Caveats}

When restructuring an expression dag there are some potential pitfalls one should be aware of. Changing the structure of an expression dag leads to a change in evaluation order, which may in turn influence the performance. Other hurdles are even more subtle, since they result from implementation details of the underlying bigfloat arithmetic. We show examples, where this leads to problems for balancing. However, the caveats are not restricted to balancing, but apply to restructuring attempts in general.

\subsection{Evaluation order}
\label{ssc:evalorder}

When evaluating a dag-based number type recursively, a slight change in expression order can have an unexpectedly high impact on the evaluation time~\cite{moerig15macis}. Balancing the dag may have a negative impact on the optimal expression order. One example where this may occur is the computation of the geometric sum \( \sum_{i=0}^n r^i \) with $r < 1$.

\begin{lstlisting}[escapechar=\%]
template <class NT> void geometric_sum(const int n, const long %\p%){
  NT r = sqrt(NT(13)/NT(64));
  NT ri = NT(1); NT s = ri; 
  for (int i=0; i<n; ++i){
    ri *= r;
    s  += ri;
  }
  s.guarantee_absolute_error_two_to(%\p%);
}
\end{lstlisting}

We call the multiplication node $m_i$ resulting from the $i$-th multiplication \emph{deeper} than the node $m_j$ resulting from the $j$-th multiplication if $i<j$ and \emph{shallower} if $j<i$. If $m_j$ is shallower than $m_i$ then $m_j$ is an ancestor of $m_i$ in the expression dag. When balancing the expression dag the accuracy needed at the deeper multiplication nodes decreases, while the accuracy needed at the shallower nodes increases. Since in \texttt{geometric\_sum} the shallower multiplication nodes depend on the deeper ones, the balancing actually increases the final accuracy needed at the deeper multiplication nodes by an amount logarithmic in the total number of additions. To make things worse, the deeper nodes are still evaluated first (with low precision) and therefore need to be recursively re-evaluated for every shallower multiplication node, leading to a quadratic number of evaluations.

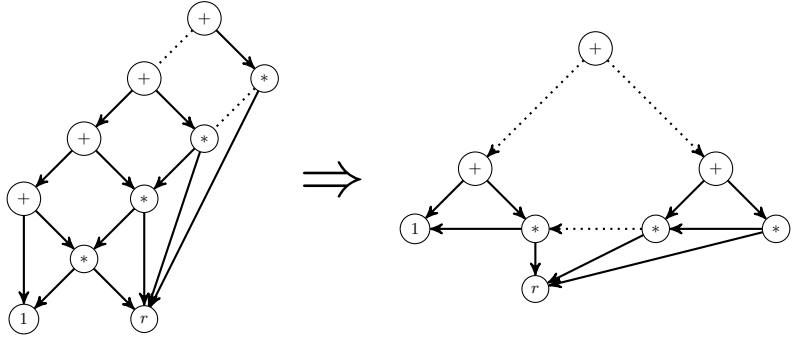
\begin{figure}
\centering
\begin{tikzpicture}
\begin{scope}[scale=0.4,shift={(-13,1)}]
  \SetGraphUnit{2} 
  \tikzstyle{VertexStyle} = [scale=0.7,circle,draw]

   \Vertex[Math,L=+]{P4} 
   \SOWE[Math,L=+](P4){P3} \SOEA[Math,L=*](P4){M4}
   \SOWE[Math,L=+](P3){P2} \SOWE[Math,L=*](M4){M3} 
   \SOWE[Math,L=+](P2){P1} \SOWE[Math,L=*](M3){M2} 
   \SOWE[Math,L=*](M2){M1} 
   \SOWE[Math,L=1](M1){ONE} \SOEA[Math,L=r](M1){R}
   
  \tikzstyle{EdgeStyle}   = [->,>=stealth']
   \Edges(P3,P2,P1,ONE) \Edges(P1,M1) \Edges(P2,M2) \Edges(P3,M3,M2,M1,ONE) \Edges(P4,M4) \Edges(M4,R) \Edges (M3,R) \Edges(M2,R) \Edges(M1,R)
  \tikzstyle{EdgeStyle}   = [dotted]
   \Edges(P4,P3) \Edges(M4,M3)
\end{scope}
\begin{scope}[shift={(-3.5,-1.8)}]
\node{\Huge $\Rightarrow$};
\end{scope}
\begin{scope}[scale=0.4]
  \SetGraphUnit{2} 
  \tikzstyle{VertexStyle} = [scale=0.7,circle,draw]

   \Vertex[Math,L=+]{P0}
   
  \tikzstyle{VertexStyle} = []
   \SOWE[NoLabel](P0){E1} \SOEA[NoLabel](P0){E2}   
   
  \tikzstyle{VertexStyle} = [scale=0.7,circle,draw]
   \SOWE[Math,L=+](E1){P1} \SOEA[Math,L=+](E2){P2}
   \SOWE[Math,L=1](P1){ONE}
   \SOEA[Math,L=*](P1){M1} \SOWE[Math,L=*](P2){M2} \SOEA[Math,L=*](P2){M3}
   \SO[Math,L=r](M1){R}

  \tikzstyle{EdgeStyle}   = [->,>=stealth']
   \Edges(M3,M2) \Edges(M1,ONE) \Edges(M3,R) \Edges(M2,R) \Edges(M1,R) \Edges(P1,ONE) \Edges(P1,M1) \Edges(P2,M2) \Edges(P2,M3)
  \tikzstyle{EdgeStyle}   = [->,>=stealth',dotted]
   \Edges(P0,P1) \Edges(P0,P2) \Edges(M2,M1)
\end{scope}
\end{tikzpicture}
\caption{Expression dags for \texttt{geometric\_sum} before and after balancing. After balancing, all multiplication nodes are on the same level, with the deeper ones evaluated first, inducing a quadratic number of evaluation steps.}
\end{figure}

Note, that this does not happen for the linear computation order if we assume the following increase in accuracy (cf.\ Section~\ref{sec:theory}):
\begin{itemize}
	\item To evaluate $z=x+y$ with accuracy $\p$, both $x$ and $y$ must be accurate up to $\p+2$ digits.
	\item To evaluate $z=x*y$ with accuracy $\p$, $x$ must be accurate up to $\p+2+\lceil\log|y|\rceil$ and $y$ must be accurate up to $\p+2+\lceil\log|x|\rceil$ digits.
\end{itemize} 
Since for $r < 1$ also $r^i \leq 1$ the increase in accuracy is the same for addition and multiplication. Therefore with linear computation order the multiplication nodes do not need to be re-evaluated after their initial evaluation. If $r>1$ the linear dag and the balanced dag show similar behavior.\footnote{\RealAlgebraic{} usually overestimates the exponent by one, therefore in our tests $r$ is chosen to be smaller than $0.5$.}

To avoid extensive recomputations, we can compute a topological order and determine the final accuracy needed at each node before recomputing it \cite{moerig15macis}. We implement this strategy and compare it with recursive evaluation.
The standard recursive evaluation procedure essentially works as depicted in Algorithm~\ref{alg:recursiveevaluation}. At each node the needed accuracy of its children is ensured and the value at this node gets recomputed. Nodes can get recomputed several times if they have more than one parent.

\begin{algorithm}[h]
 \KwData{requested accuracy $\p$}
 \If{error is larger than $2^{-\p}$}{
   compute needed accuracy for children\\
   recurse on children with their respective accuracy\\
   recompute
 }
\caption{Evaluating an expression dag by recursively increasing the accuracy of the children before recomputing the current operation.}
\label{alg:recursiveevaluation}
\end{algorithm}

When evaluating topologically we determine a topological order for all inexact nodes and compute the maximum accuracy needed for those nodes. Afterwards we recompute the nodes with their maximum accuracy (cf.\ Algorithm~\ref{alg:topologicalevaluation}). By following this procedure we can guarantee that no node is recomputed more than once during one evaluation of the expression dag.

\begin{algorithm}[htb]
 \KwData{requested accuracy $\p$}
 \If{error is larger than $2^{-\p}$}{
   $top =$ all inexact nodes in topological order\\
   \For{$i=1$ \textbf{to} $|top|$}{
     update the required error for the children of $top[i]$
   }
   \For{$i=|top|$ \textbf{downto} $1$}{
     \If{$top[i].error > top[i].requested\_error$}{
       recompute $top[i]$
     }
   }
 }
\caption{Evaluating an expression dag by finding a topological order and determining the maximum accuracy needed at each node before recomputing them.}
\label{alg:topologicalevaluation}
\end{algorithm}

We execute the geometric sum experiment with the four balancing strategies from before. Furthermore for each of these strategies we evaluate either recursively (r) or in topological order (t).

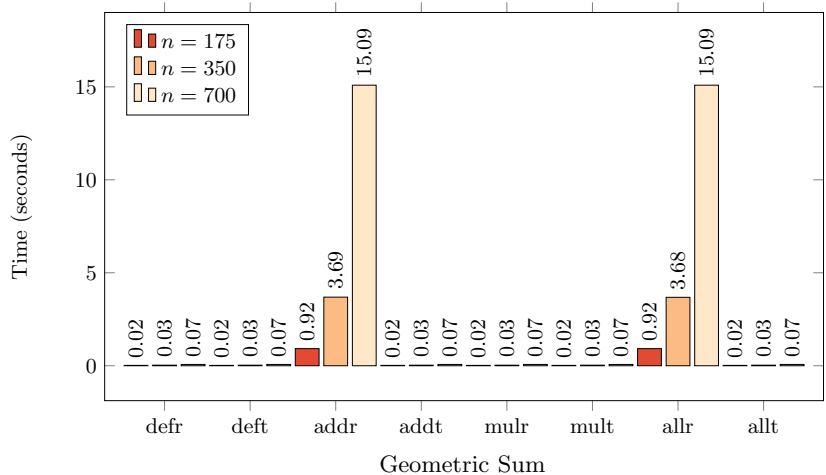
\begin{figure}[h]
\centering
\captionsetup[subfigure]{labelformat=empty}
\subfloat[Geometric Sum]{
\hspace{-40pt}
\begin{tikzpicture}[scale=0.9]
\begin{axis}[
	symbolic x coords={defr,deft,addr,addt,mulr,mult,allr,allt},
	xtick=data,
	ylabel=Time (seconds),
	ymax = 19,
    width  = \textwidth,
    height  = 0.6\textwidth,
    ybar,
    legend pos=north west,
    nodes near coords,
    every node near coord/.append style={rotate=90, anchor=west, 
    									 /pgf/number format/precision=4}
]

\addplot[fill=SEQCOL1] coordinates {(defr,0.02) (deft,0.02) (addr,0.92) (addt,0.02) (mulr,0.02) (mult,0.02) (allr,0.92) (allt,0.02) };
\addplot[fill=SEQCOL2] coordinates {(defr,0.03) (deft,0.03) (addr,3.69) (addt,0.03) (mulr,0.03) (mult,0.03) (allr,3.68) (allt,0.03) };
\addplot[fill=SEQCOL3] coordinates {(defr,0.07) (deft,0.07) (addr,15.09) (addt,0.07) (mulr,0.07) (mult,0.07) (allr,15.09) (allt,0.07) };
	
\legend{$n=175$,$n=350$,$n=700$}
\end{axis}
\end{tikzpicture}
}
\caption{Balancing additions leads to a massive increase in running time for \texttt{geometric\_sum} with $\p=50000$ by creating a bad evaluation order. Topological evaluation solves the problem.}
\label{fig:geosum}
\end{figure}

As the results in Figure~\ref{fig:geosum} show, balancing the expression dag destroys a favorable evaluation order when computing the geometric sum. Switching to a topological evaluation order negates this effect. Note that the performance loss due to the logarithmic increase of precision in the balanced case is too small to show in the measurements.

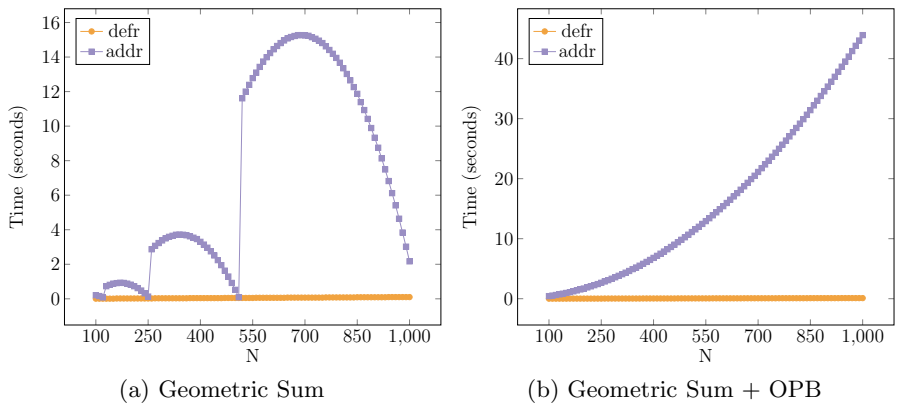
\begin{figure}[H]
\centering
\subfloat[Geometric Sum]{
\begin{tikzpicture}[scale=0.5]
\begin{axis}[
	xtick={100,250,...,1000},
	xlabel=N,
	ylabel=Time (seconds),
    label style={font=\Large},
    tick label style={font=\Large},
    legend style={font=\Large},
    legend pos=north west,
    width  = 0.95*\textwidth
]
\addplot[QCOL1,mark=*] coordinates { (100,0.01) (110,0.01) (120,0.01) (130,0.01) (140,0.01) (150,0.01) (160,0.02) (170,0.02) (180,0.02) (190,0.02) (200,0.02) (210,0.02) (220,0.02) (230,0.02) (240,0.02) (250,0.02) (260,0.02) (270,0.03) (280,0.03) (290,0.03) (300,0.03) (310,0.03) (320,0.03) (330,0.03) (340,0.03) (350,0.03) (360,0.03) (370,0.04) (380,0.04) (390,0.04) (400,0.04) (410,0.04) (420,0.04) (430,0.04) (440,0.04) (450,0.04) (460,0.04) (470,0.05) (480,0.05) (490,0.05) (500,0.05) (510,0.05) (520,0.05) (530,0.05) (540,0.05) (550,0.05) (560,0.06) (570,0.06) (580,0.06) (590,0.06) (600,0.06) (610,0.06) (620,0.06) (630,0.06) (640,0.06) (650,0.07) (660,0.07) (670,0.07) (680,0.07) (690,0.07) (700,0.07) (710,0.07) (720,0.07) (730,0.07) (740,0.07) (750,0.07) (760,0.08) (770,0.08) (780,0.08) (790,0.08) (800,0.08) (810,0.08) (820,0.08) (830,0.08) (840,0.08) (850,0.09) (860,0.09) (870,0.09) (880,0.09) (890,0.09) (900,0.09) (910,0.09) (920,0.09) (930,0.09) (940,0.1) (950,0.1) (960,0.1) (970,0.1) (980,0.1) (990,0.1) (1000,0.1)};

\addplot[QCOL2,mark=square*] coordinates { (100,0.21) (110,0.16) (120,0.08) (130,0.73) (140,0.81) (150,0.88) (160,0.91) (170,0.93) (180,0.92) (190,0.88) (200,0.82) (210,0.73) (220,0.62) (230,0.48) (240,0.32) (250,0.13) (260,2.87) (270,3.07) (280,3.23) (290,3.38) (300,3.5) (310,3.59) (320,3.65) (330,3.71) (340,3.72) (350,3.71) (360,3.68) (370,3.62) (380,3.54) (390,3.46) (400,3.3) (410,3.14) (420,2.96) (430,2.75) (440,2.51) (450,2.24) (460,1.95) (470,1.63) (480,1.29) (490,0.92) (500,0.52) (510,0.09) (520,11.61) (530,12) (540,12.39) (550,12.78) (560,13.1) (570,13.43) (580,13.74) (590,13.99) (600,14.23) (610,14.45) (620,14.64) (630,14.81) (640,14.96) (650,15.08) (660,15.16) (670,15.22) (680,15.27) (690,15.27) (700,15.25) (710,15.23) (720,15.15) (730,15.06) (740,14.95) (750,14.81) (760,14.62) (770,14.41) (780,14.19) (790,13.95) (800,13.67) (810,13.35) (820,13.02) (830,12.65) (840,12.26) (850,11.87) (860,11.39) (870,10.93) (880,10.42) (890,9.89) (900,9.33) (910,8.75) (920,8.13) (930,7.49) (940,6.82) (950,6.12) (960,5.41) (970,4.64) (980,3.84) (990,3.02) (1000,2.17)};
	
\legend{defr,addr}
\end{axis}
\end{tikzpicture}
}
\subfloat[Geometric Sum + OPB]{
\begin{tikzpicture}[scale=0.5]
\begin{axis}[
	xtick={100,250,...,1000},
	xlabel=N,
	ylabel=Time (seconds),
    label style={font=\Large},
    tick label style={font=\Large},
    legend style={font=\Large},
    legend pos=north west,
    width  = 0.95*\textwidth
]

\addplot[QCOL1,mark=*] coordinates { (100,0.01) (110,0.01) (120,0.01) (130,0.01) (140,0.01) (150,0.01) (160,0.02) (170,0.02) (180,0.02) (190,0.02) (200,0.02) (210,0.02) (220,0.02) (230,0.02) (240,0.02) (250,0.02) (260,0.03) (270,0.03) (280,0.03) (290,0.03) (300,0.03) (310,0.03) (320,0.03) (330,0.03) (340,0.03) (350,0.03) (360,0.03) (370,0.04) (380,0.04) (390,0.04) (400,0.04) (410,0.04) (420,0.04) (430,0.04) (440,0.04) (450,0.04) (460,0.04) (470,0.05) (480,0.05) (490,0.05) (500,0.05) (510,0.05) (520,0.05) (530,0.05) (540,0.05) (550,0.05) (560,0.05) (570,0.06) (580,0.06) (590,0.06) (600,0.06) (610,0.06) (620,0.06) (630,0.06) (640,0.06) (650,0.06) (660,0.07) (670,0.07) (680,0.07) (690,0.07) (700,0.07) (710,0.07) (720,0.07) (730,0.07) (740,0.07) (750,0.07) (760,0.08) (770,0.08) (780,0.08) (790,0.08) (800,0.08) (810,0.08) (820,0.08) (830,0.08) (840,0.08) (850,0.09) (860,0.09) (870,0.09) (880,0.09) (890,0.09) (900,0.09) (910,0.09) (920,0.09) (930,0.09) (940,0.1) (950,0.1) (960,0.1) (970,0.1) (980,0.1) (990,0.1) (1000,0.1)};

\addplot[QCOL2,mark=square*] coordinates { (100,0.43) (110,0.52) (120,0.62) (130,0.72) (140,0.84) (150,0.96) (160,1.09) (170,1.23) (180,1.38) (190,1.54) (200,1.7) (210,1.88) (220,2.06) (230,2.25) (240,2.44) (250,2.65) (260,2.87) (270,3.09) (280,3.33) (290,3.57) (300,3.82) (310,4.08) (320,4.34) (330,4.62) (340,4.93) (350,5.2) (360,5.5) (370,5.81) (380,6.13) (390,6.46) (400,6.8) (410,7.14) (420,7.5) (430,7.86) (440,8.24) (450,8.62) (460,9.01) (470,9.42) (480,9.82) (490,10.24) (500,10.67) (510,11.11) (520,11.55) (530,11.99) (540,12.46) (550,12.93) (560,13.41) (570,13.9) (580,14.39) (590,14.91) (600,15.42) (610,15.95) (620,16.5) (630,17.04) (640,17.59) (650,18.16) (660,18.73) (670,19.31) (680,19.93) (690,20.49) (700,21.12) (710,21.74) (720,22.37) (730,23.01) (740,23.66) (750,24.32) (760,24.98) (770,25.67) (780,26.35) (790,27.04) (800,27.74) (810,28.45) (820,29.17) (830,29.91) (840,30.66) (850,31.42) (860,32.21) (870,32.99) (880,33.78) (890,34.55) (900,35.35) (910,36.16) (920,36.98) (930,37.82) (940,38.66) (950,39.52) (960,40.35) (970,41.26) (980,42.14) (990,43.02) (1000,43.92)};

\legend{defr,addr}
\end{axis}
\end{tikzpicture}
}
\caption{Comparison of the behavior of \texttt{geometric\_sum} for increasing $n$ with the balancing procedure from algorithm~\ref{alg:balanceoperation} and with order preserving balancing (OPB). The original procedure leads to jumps in running time, while order preserving balancing produces the expected quadratic behavior ($\p=50000$, averaged over $5$ runs each).}
\label{fig:geosumseries}
\end{figure}

The values for $n$ have been chosen to show spikes in the running time. As pointed out in Section~\ref{sec:implementation} our balancing algorithm does not necessarily preserve the order of the operands. If the shallowest multiplication node is evaluated first, this leads to an optimal evaluation order. Figure~\ref{fig:geosumseries} compares our implementation with an order-preserving balancing strategy.\footnote{This is implemented by inserting dummy nodes up to the next power of two.}

The algorithm we use to build a balanced tree results in large jumps when stepping from $2^k-1$ to $2^k$ operands ($k\in\mathbb{N}$). With $2^k-1$ operands the previously rightmost operand, i.e.\ the shallowest multiplication node, becomes the leftmost operand in the balanced tree and therefore the evaluation order is optimal. With $2^k$ operands the previous operand order is preserved by the algorithm and is therefore the worst possible.
If preserving order is enforced, the quadratic increase in running time is evident.

\subsection{Operands matter}

In some cases balancing can destroy a favorable dag structure independently from the evaluation order. We compute the telescoping product $\prod_{i=1}^{n-1}\frac{i+1}{i}$ through the following algorithm.

\begin{lstlisting}[escapechar=\%]
template <class NT>
void telescoping_product(const int n, const long %\p%){
  NT prod = NT(1);
  for (int i = 1; i < n; ++i) {
    prod *= NT(i+1)/NT(i);
  }
  prod.guarantee_absolute_error_two_to(%\p%);
}
\end{lstlisting}

In the experimental results shown in Figure~\ref{sfig:telpi1}, a performance decrease due to balancing is evident, which also cannot be corrected through a change in evaluation order. The reason for this effect is that the naive order enables the bigfloat arithmetic to make use of eliminating factors. Bigfloat multiplications involving integers\footnote{Or integers divided by a power of two.} are less expensive\footnote{This behavior was confirmed with both \texttt{mpfr} and \texttt{leda} bigfloats.}. In the original expression order the result of each multiplication is an integer and can be determined as such. Therefore, although significantly reducing the average need of accuracy, balancing has a negative effect on the performance.

\begin{figure}[h]
\centering
\subfloat[Telescoping Product $i\geq1$]{
\begin{tikzpicture}[scale=0.5]
\begin{axis}[
	symbolic x coords={defr,deft,mulr,mult},
	xtick=data,
	enlarge x limits=0.2,
	ylabel=Time (seconds),
	ymin = 0,
	ymax = 4.5,
    width  = 0.95*\textwidth,
    bar width = 0.5cm,
    ybar,
    label style={font=\Large},
    tick label style={font=\Large},
    legend style={font=\Large,at={(0.02,0.88)},anchor=west},
    nodes near coords,
    every node near coord/.append style={rotate=90, anchor=west, 
    									 /pgf/number format/precision=4,font=\Large}
]

\addplot[fill=SEQCOL1] coordinates {(defr,0.10) (deft,0.10) (mulr,0.85) (mult,0.85) };
\addplot[fill=SEQCOL2] coordinates {(defr,0.23) (deft,0.23) (mulr,1.70) (mult,1.71) };
\addplot[fill=SEQCOL3] coordinates {(defr,0.54) (deft,0.55) (mulr,3.42) (mult,3.43) };
	
\legend{$n=5000$,$n=10000$,$n=20000$}
\end{axis}
\end{tikzpicture}
\label{sfig:telpi1}
}
\subfloat[Telescoping Product $i\geq3$]{
\begin{tikzpicture}[scale=0.5]
\begin{axis}[
	symbolic x coords={defr,deft,mulr,mult},
	xtick=data,
	enlarge x limits=0.2,
	ylabel=Time (seconds),
	ymin = 0,
	ymax = 6.5,
    width  = 0.95*\textwidth,
    bar width = 0.5cm,
    ybar,
    label style={font=\Large},
    tick label style={font=\Large},
    legend style={font=\Large},
    nodes near coords,
    every node near coord/.append style={rotate=90, anchor=west, 
    									 /pgf/number format/precision=4,font=\Large}
]
\addplot[fill=SEQCOL1] coordinates {(defr,0.77) (deft,0.78) (mulr,0.85) (mult,0.86) };
\addplot[fill=SEQCOL2] coordinates {(defr,1.92) (deft,1.92) (mulr,1.71) (mult,1.72) };
\addplot[fill=SEQCOL3] coordinates {(defr,5.50) (deft,5.51) (mulr,3.42) (mult,3.43) };
	
\legend{$n=5000$,$n=10000$,$n=20000$}
\end{axis}
\end{tikzpicture}
\label{sfig:telpi3}
}
\caption{Performance of \texttt{telescoping\_product} starting with $i=1$ or $i=3$ before and after balancing multiplications ($\p=50000$). Balancing destroys a favorable order of the operands, which cannot be corrected for by switching to topological evaluation. For $i=3$ the original order was less favorable and the performance gain of balancing outweighs the loss for larger $n$.}
\label{fig:telprod}
\end{figure}
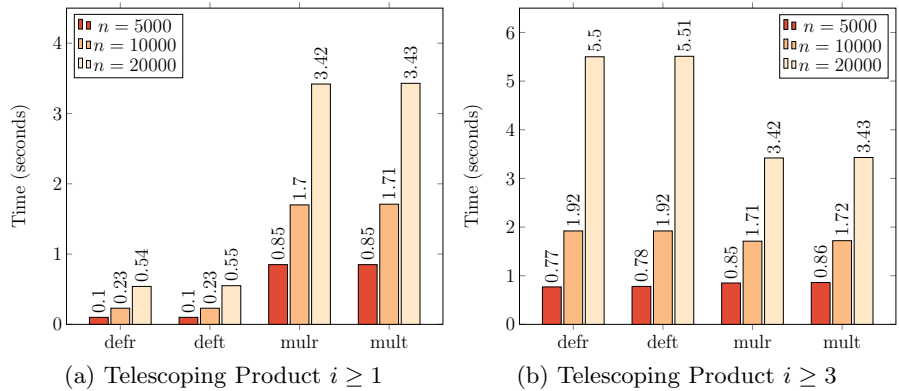

If the product is computed starting with $i=3$ only every third subexpression evaluates to an integer. While there are still some favorable structures getting disrupted by balancing the expression dag, the benefit of balancing surpasses the loss as the number of operands increases (cf.\ Figure~\ref{sfig:telpi3}).
The effect vanishes if the product is computed in reverse order as depicted in the following algorithm. 

\begin{lstlisting}[escapechar=\%]
template <class NT>
void telescoping_product_reverse(const int n, const long %\p%){
  NT prod = NT(1);
  for (int i = n-1; i >= 1; --i) {
    prod *= NT(i+1)/NT(i);
  }
  prod.guarantee_absolute_error_two_to(%\p%);
}
\end{lstlisting}

By this, none of the subexpressions involved evaluates to an integer and only a logarithmic amount of subexpressions evaluates to an integer divided by a power of two. The results of the experiment for the reverse case are shown in Figure~\ref{sfig:telprodreverse}. Now balancing has the expected positive effect on the overall performance. Note that, as expected, the forward loop starting with $i=3$ without balancing takes approximately two third of the time of the reverse case.

\begin{figure}[htb]
\ContinuedFloat
\captionsetup{list=off,format=cont}
\captionsetup[subfigure]{format=plain}
\centering
\subfloat[Telescoping Product Reverse]{
\begin{tikzpicture}[scale=0.5]
\begin{axis}[
	symbolic x coords={defr,deft,mulr,mult},
	xtick=data,
	enlarge x limits=0.2,
	ylabel=Time (seconds),
	ymax = 9.5,
    width  = 0.95*\textwidth,
    bar width = 0.5cm,
    ybar,
    label style={font=\Large},
    tick label style={font=\Large},
    legend style={font=\Large},
    nodes near coords,
    every node near coord/.append style={rotate=90, anchor=west, 
    									 /pgf/number format/precision=4,font=\Large}
]

\addplot[fill=SEQCOL1] coordinates {(defr,1.10) (deft,1.10) (mulr,0.85) (mult,0.85) };
\addplot[fill=SEQCOL2] coordinates {(defr,2.75) (deft,2.75) (mulr,1.70) (mult,1.71) };
\addplot[fill=SEQCOL3] coordinates {(defr,7.91) (deft,7.92) (mulr,3.41) (mult,3.43) };
	
\legend{$n=5000$,$n=10000$,$n=20000$}
\end{axis}
\end{tikzpicture}
\label{sfig:telprodreverse}
}
\caption{Performance of \texttt{telescoping\_product\_reverse} before and after balancing multiplications ($\p=50000$). No favorable order is destroyed in contrast to \texttt{telescoping\_product} (cf.\ Figure~\ref{sfig:telpi1}) and balancing shows the expected net benefit.}
\label{fig:telprodreverse}
\end{figure}
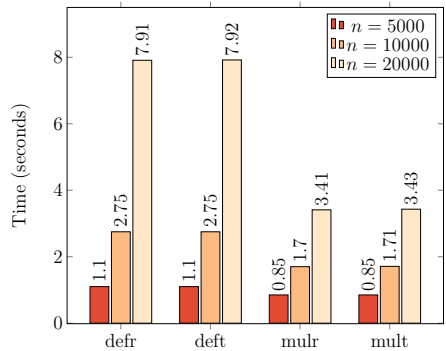

\subsection{Overhead}

In all tests, except the telescoping product test with $i=1$, the overhead for the balancing procedure as well as for the topological sorting was (usually much) less than $0.5\%$ of the final running time.
The running time of \texttt{telescoping\_product} is unusually small compared to its number of operations, therefore the relative overhead of additional computations is higher. In this case the overhead amounts to less than $2\%$ for balancing and less than $3\%$ for topological sorting.

\section{Conclusion}

Balancing additions and multiplications in an expression dag can significantly reduce the computation time needed as demonstrated by the sum-of-square-roots test and the binomial coefficient test. The experimental data indicates that the overhead due to the balancing algorithm is small compared to the cost of the bigfloat operations.
Balancing may cause changes in the evaluation order that lead to increased running time. Those issues can partially be addressed by switching to a topological evaluation, which can be done with small overhead as well.

We conclude that it is useful to provide a number type supporting balancing of expression dags in combination with topological evaluation. The use of this number type should be considered whenever an algorithm performs a large number of consecutive additions or multiplications. Switching a number type is usually less time-consuming than a deep analysis and adjustment of the used algorithm.

\section{Future work}

In this paper we restricted restructuring of the dag to balancing additions and multiplications. Performance increase due to further restructuring is imaginable. Subtractions could easily be included in the balancing process by treating them like an addition and a negation and propagating the negations to the operands. It may also be useful to incorporate divisions into the multiplication balancing process. Since inversions are much more expensive than negations, it seems not feasible to replace them by a multiplication and an inversion. Instead a promising strategy might be to reduce the number of divisions by raising them to the root.

Balancing an expression dag makes its nodes more independent and therefore makes it more accessible for parallelization. Further restructuring with the goal of faster parallelization, e.g.\ expanding products, might be profitable.

\bibliography{lit}
\bibliographystyle{splncs03}
\end{document}